\documentclass[letterpaper, 10 pt, conference]{ieeeconf}  % Comment this line out if you need a4paper

\IEEEoverridecommandlockouts                          

\usepackage{graphics} % for pdf, bitmapped graphics files
\usepackage{epsfig} % for postscript graphics files
\usepackage{mathptmx} % assumes new font selection scheme installed
\usepackage{times} % assumes new font selection scheme installed
\usepackage{amsmath} % assumes amsmath package installed
\usepackage{amssymb}  % assumes amsmath package installed
\usepackage{cancel}
\usepackage{float}
\usepackage{diagbox}
\usepackage{makecell}  % Include the makecell package
\usepackage{color}

\usepackage{graphicx}
\usepackage{subcaption}
\usepackage{lipsum} % For dummy text, remove in actual document

\newcommand{\R}{{\mathbb{R}}}
\newcommand{\N}{{\mathbb{N}}}

\newtheorem{definition}{Definition}
\newtheorem{theorem}{Theorem}
\newtheorem{remark}{Remark}
\newcommand{\nex}{\mathord{\bigcirc}}
\title{\LARGE \bf
Maximally Resilient Controllers under Temporal Logic Specifications
}

\author{Youssef Ait Si$^{1}$, Ratnangshu Das$^{2}$, Negar Monir$^{3}$,  \\  Sadegh Soudjani$^{4}$, Pushpak Jagtap$^{2}$, and Adnane Saoud$^{1}$% <-this % stops a space
\thanks{$^{1}$ College of Computing, University Mohammed VI Polytechnic, Benguerir, Morocco. {\tt\small \{youssef.aitsi, adnane.saoud\}@um6p.ma}
}%
\thanks{$^{2}$The Robert Bosch Centre for Cyber-Physical Systems, IISc, Bangalore, India {\tt\small \{ratnangshud,pushpak\}@iisc.ac.in}}%
\thanks{$^{3}$ Newcastle University, Newcastle upon Tyne, United Kingdom {\tt\small S.Seyedmonir2@newcastle.ac.uk}
        }%
\thanks{$^{4}$ Max Planck Institute for Software Systems, Kaiserslautern, Germany {\tt\small sadegh@mpi-sws.org}
        }%
}

\begin{document}

\maketitle
\thispagestyle{empty}
\pagestyle{empty}

%%%%%%%%%%%%%%%%%%%%%%%%%%%%%%%%%%%%%%%%%%%%%%%%%%%%%%%%%%%%%%%%%%%%%%%%%%%%%%%%
\begin{abstract}
In this paper, we consider the notion of resilience of a dynamical system, defined by the maximum disturbance a controlled dynamical system can withstand while satisfying given temporal logic specifications. Given a dynamical system and a specification, the objective is to synthesize the controller such that the closed-loop system satisfies this specification while maximizing its resilience. The problem is formulated as a robust optimization program where the objective is to compute the maximum resilience while simultaneously synthesizing the corresponding controller parameters. For linear systems and linear controllers, exact solutions are provided for the class of time-varying polytopic specifications. For the case of nonlinear systems, nonlinear controllers and more general specifications, we leverage tools from the scenario optimization approach, offering a probabilistic guarantee of the solution as well as computational feasibility. Different case studies are presented to illustrate the theoretical results. 
\end{abstract}

%%%%%%%%%%%%%%%%%%%%%%%%%%%%%%%%%%%%%%%%%%%%%%%%%%%%%%%%%%%%%%%%%%%%%%%%%%%%%%%%
\section{INTRODUCTION}
In real-world control applications, disturbances, model uncertainties, and environmental variations are inevitable. Autonomous vehicles must maintain safe operation despite road conditions and sensor noise \cite{vargas2021overview}, robotic manipulators need to perform tasks accurately in unpredictable settings \cite{kolhe2013robust}, and power grids must maintain stability even with fluctuating demands and faults \cite{zhang2021grid}. Traditional robust control methods focus on ensuring that a system remains within predefined performance under bounded disturbances. However, these approaches primarily emphasize maintaining safety and performance rather than quantifying the maximum disturbance a system tolerate before it fails to meet its intended specifications. This gap motivates the need for a resilience metric a way to quantify the maximum disturbance a system can withstand while still satisfying its required specifications.

Resilience is particularly important when dealing with temporal specifications \cite{fainekos2009robustness} to enforce complex behavioral constraints over time, including safety and reachability with precise timings. Temporal logic has been widely used in control design \cite{kloetzer2008fully} to impose spatial and temporal constraints on the system behavior, such as ``reach a target within a time limit" or ``avoid obstacles at all times." Although existing robust control approaches ensure specification satisfaction under disturbances \cite{sadraddini2015robust, haesaert2020robust, das2025spatiotemporal}, they do not directly focus on maximizing resilience. 

Several notions of resilience have been explored in the control systems literature~\cite{AitS2025, rieger2009resilient}. The authors in \cite{rieger2009resilient} define resilience as the system's ability to maintain state awareness and functionality in response to disturbances. The authors in \cite{zhu2011robust, monir2025} present a holistic theoretical model for robust and resilient control in power systems, focusing on voltage regulation under sudden faults or attacks. More recently, \cite{chen2023stl} have introduced a resilience framework that combines recoverability (how quickly a system recovers from a failure) and durability (how well it avoids failures after recovery) and \cite{monir2025AGC} demonstrate how resilience can be used to compute feasible assume–guarantee contracts. However, most of these works consider resilience in terms of restoration of a system under unexpected extreme and rare events, and do not directly quantify how much disturbance a system can handle. {The concept of resilience as the maximum disturbance a dynamical system can tolerate while ensuring temporal logic satisfaction that was formally introduced for non controlled systems in} \cite{Saoud2023, saoud2024}. {In this paper, we extend this concept for controlled systems where} the key challenge lies in synthesizing controllers that not only ensure satisfaction of specifications under nominal conditions but also maximize resilience against disturbances. This paper builds on this idea by formulating resilient control synthesis as a robust optimization problem, where the goal is to design an optimal controller that ensures the system satisfies a finite horizon specification enforcing the system trajectories to maintain a defined behavior and, at the same time, find the largest possible set of admissible disturbances. Our results generalize the work in \cite{saoud2024} from non controlled to controlled discrete time systems by introducing a new resilience metric. In addition, we provide tractable algorithms to compute the resilience metric as well the optimal controller for linear and nonlinear discrete time dynamical system.

The contributions developed in this paper are as follows: defining the resilience metric for a controlled dynamical system as a robust optimization problem. In the particular case of a linear system with a linear controller, we translate this robust optimization problem defined by the resilience metric into a tractable optimization problem with polynomial constraints, yielding exact solutions. In the general case of nonlinear systems and nonlinear controllers and a general finite-horizon specification, we leverage scenario optimization techniques to compute an approximation of the resilience metric with probabilistic guarantees. Moreover, we defined a general framework for finite specifications, where finite-horizon safety and exact time reachability are considered particular cases.

The remainder of this paper is structured as follows. Section~\ref{sec:2} formally defines the controlled system dynamics, temporal specifications, and the resilience metric. Section~\ref{sec:3} presents results for computing exact resilience for linear systems with linear controllers. Section~\ref{scenario_section} extends the analysis to nonlinear systems and nonlinear controllers. Section~\ref{sec:4} provides two case studies, demonstrating the effectiveness of our approach. The proofs of theorems are relegated to the appendix.

\section{Preliminaries and Problem Formulation}

\label{sec:2}
\textbf{Notations:}
The symbols $ \mathbb{N} $, $ \mathbb{N}_{\geq 0} $, $ \R$, and $\R_{\geq 0}$ denote the set of positive integers, nonnegative integers, real, and non-negative real numbers, respectively. We use $\mathbb{R}^{n \times m}$ to denote the space of real matrices with $n$ rows and $m$ columns. For a matrix $A \in \mathbb{R}^{n \times m}$, $A^T$ represents the transpose of $A$. For a vector $x \in \mathbb{R}^n$, we use $\|x\|$ and $\|x\|_{\infty}$ to denote the Euclidean and infinity norm, respectively. We use $\mathbb{I}_n$ to denote the identity matrix of size $n \times n$ and similarly $0_n$ and $0_{n\times m}$ represent the zero matrix of size $n \times n$ and $n\times m$ respectively.
Given $x \in \mathbb{R}^n$ and $\varepsilon \geq 0$, $\Omega_{\varepsilon}(x)=\left\{z \in \mathbb{R}^n \mid\|z-x\|_{\infty} \leq \varepsilon\right\}$ and $\mathcal{B}_{\varepsilon}(x)=\left\{z \in \mathbb{R}^n \mid\|z-x\| \leq \varepsilon\right\}$. The combination of \(k \in  \N\) items chosen from \(n \in \N\) distinct items is given by the formula $\binom{n}{k} = \frac{n!}{k!(n-k)!}$, where \(n!\) is the factorial of \(n\).

\subsection{Discrete-time Dynamical Systems}
\label{system definition}
A discrete-time system can be defined as a tuple $\Sigma = (X,U,D,f)$, where
$X\subset \mathbb{R}^n$ is the state space, $U\subset \mathbb{R}^m$ is the input space of the system, and
$D\subset\mathbb{R}^n$ is the disturbance space which is assumed to be a compact set and contains the origin and $f: X \times U \rightarrow X$ is a continuous map representing the system dynamics. 
The system $\Sigma$ is evolving according to the following dynamics:
\begin{equation}
\label{eqn:non linear controlled system}
x(k+1) = f(x(k) , u(k)) + d(k),\quad  k \in \mathbb{N},
\end{equation}
where $x(k) \in X$, $u(k)  \in U $, and $d(k)\in D$ represent the system state, system input, and the additive disturbance, respectively, at time $k$. 
In this work, the system $\Sigma$ is controlled by a continuous state feedback controller $\pi_\alpha$ defined by $\pi_\alpha : X \to U$ such that \( \pi_\alpha(x(k)) =  u(k)\), where $ \alpha \in \R^{d}$ represent the collection of parameters of the controllers. A simple example is a linear controller, which can be modeled by the function $\pi_\alpha(x) = \alpha_1 x + \alpha_2$, where $(\alpha_1, \alpha_2) \in \mathbb{R}^{m\times n} \times \mathbb{R}^{m}$ are two real matrices of dimension $m \times n$ and $m \times 1$, respectively. In this context, the set of parameters $\alpha$ is given by $\alpha = ( \alpha_1, \alpha_2) \in \R^{m\times n + m}$. Another notable example is polynomial controllers \cite{polynomial2013}. 
Consider the control function \(\pi_\alpha(x): \mathbb{R}^n \rightarrow \mathbb{R}^m\) defined as a polynomial of degree \(l\) given by \(\pi_\alpha(x) = \alpha \xi(x)\), where \(\xi(x) = \left[ 1, x^{[1]}, x^{[2]}, \ldots, x^{[l]} \right]\) is a vector of monomials of degree up to \(l\), with each \(x^{[i]} \in \mathbb{R}^{d(n,i)}\) containing all distinct monomials of degree \(i\) with no repeated elements. The dimension of \(x^{[i]}\) is \(d(n,i) = \binom{n+i-1}{n-1}\) and the total number of distinct monomials up to degree \(l\) is \(D(n,l) = \sum_{i=0}^{l} d(n,i) = \binom{n+l}{n}\), which is the dimension of the vector \(\xi(x)\). In this setup, we have $\alpha = [\alpha_0, \alpha_1, \ldots, \alpha_l]$ representing a vector of dimension $ m \times D(n,l)$, where $\alpha_i \in \mathbb{R}^{m \times d(n, i)}$.

\subsection{Temporal specification}
\label{Sec:specfication}
Consider the discrete time system $\Sigma$ in \eqref{eqn:non linear controlled system}. A specification $\psi \subset X^{N+1}$ is a set of admissible state sequences that defines the desired behavior of the system over a bounded time horizon $N \in \N$. This class of specifications is quite rich, and can cover specifications such as safety, reachability and more complex linear temporal logic specifications over finite traces, LTL$_f$ \cite{LTLf17}. For example, an exact-time reachability at time $M \in \{0,1,\ldots,N\}$ of a set $A \subseteq X$, which is written in LTL$_f$ as $\psi = \nex^MA$, can be formulated as $$\psi = X^M\times A \times  X^{N-M} \subseteq X^{N+1}.$$
Similarly, finite-horizon reachability of a set $A \subseteq X$ between time instances $M_1$ and $M_2$, with $0 \leq M_1 < M_2 \leq N$, which is denoted by $\psi =\lozenge^{[M_1,M_2]} A$, can be formulated as
$$\psi= \bigcup\limits_{i=M_1}^{M_2}X^{i}\times A \times  X^{N-{i}} \subseteq X^{N+1}. $$
Finally, the finite-horizon safety of a set $A \subseteq X$ between time instances $M_1$ and $M_2$, with $0 \leq M_1 < M_2 \leq N$, which is denoted $\psi =\square^{[M_1,M_2]} A$, can be formulated as
$$\psi= \bigcap\limits_{i=M_1}^{M_2}X^{i}\times A \times  X^{N-{i}} \subseteq X^{N+1}. $$
For the system $\Sigma$ in (\ref{eqn:non linear controlled system}), we use $ \xi( x, \pi_\alpha,\mathbf{d})$ to denote the state-input trajectory of the closed-loop system, over a bounded time horizon of length $N \in \N$, starting from a state $ x \in X$ under the feedback controller $\pi_\alpha$ and the disturbance input $ \mathbf{d} = (d(0),\ldots, d(N-1)) \in D^N$. We use $ \xi( x, \pi_\alpha,D)$ to denote the set of all state-input trajectories of the closed loop system, over a bounded time horizon of length $N \in \N$, starting from a state $ x \in X$ under the feedback controller $\pi_\alpha$ under all the possible disturbance inputs $ \mathbf{d} = ( d(0),\ldots, d(N-1)) \in D^N$, defined formally as follows:
\begin{equation*}
\begin{aligned}
\label{trajectories set}
    \xi(x, \pi_\alpha, D&) =  \{((x(0), u(0)), (x(1),u(1)), \dots, (x(N)) \mid x(0) = x,\\ &x(k+1) = f(x(k) , u(k)) + d(k), u(k) = \pi_\alpha(x(k)), \\&  
    \text{ for all } d(k)   \in D   \text{ with } k \in \{0, \dots, N-1\}\}.
\end{aligned}
\end{equation*}
Additionally, we denote the projection of the trajectories on the state space as $ \xi_x(x, \pi_\alpha, D) \subseteq X^{N+1}$ and the projection on the input space as $\xi_u(x,\pi_\alpha, D) \subseteq U^{N}$ for $N \in \N $. The system $\Sigma$ in (\ref{eqn:non linear controlled system}) starting from $ x \in X$ under the feedback controller $\pi_\alpha$ is said to satisfy the specification $\psi \subseteq X^{N+1}$ if $\xi_x( x,\pi_\alpha,D) \subseteq \psi$.

In the rest of the paper, we focus on disturbance sets defined by a ball centered at zero with respect to the infinity norm, denoted as \( D := \Omega_{\varepsilon}(0) \). For simplicity, we use the shorthand notation $\xi(x, \pi_\alpha, \varepsilon) := \xi(x, \pi_\alpha, \Omega_{\varepsilon}(0)),$ and similar notation for \( \xi_x \) and \( \xi_u \).

\subsection{Resilience metric}
In this section, we define the resilience metric for a controlled discrete-time system as in \eqref{eqn:non linear controlled system} with the set of disturbances given by a ball centered at zero: $D := \Omega_{\varepsilon}(0)$. This definition extends the concept originally introduced for non-controlled systems in \cite{saoud2024}. 
\begin{definition}
\label{def:controlled_resilience}
Consider the discrete-time dynamical system $\Sigma$ in \eqref{eqn:non linear controlled system}, a specification $\psi \subseteq X^{N+1}$, a controller $\pi_\alpha$ as in Section~\ref{system definition} and a point $x \in X$. We define the {\it resilience} of the system $\Sigma$ with respect to the initial condition $x$ and the specification $\psi$ as a function $g_{\psi}:X\rightarrow \mathbb{R}_{\ge 0}\cup\{+\infty\}$ with
\begin{equation}
\label{eq:resilience}\hspace{-0.2em} 
g_\psi(x) = 
\begin{cases}
\begin{aligned}
&\sup_{\varepsilon \geq 0, \alpha \in \R^d}\big\{\varepsilon\ge 0\,|\,\xi_x(x, \pi_\alpha,\varepsilon)\subseteq \psi \big\}, \\ 
& \hspace{2.4cm} \text{ if } \exists \alpha \in \R^d , \xi_x(x, \pi_\alpha,0) \in\psi\\
&0,  \hspace{2cm} \text{ if } \forall \alpha \in \R^d , \xi_x( x, \pi_\alpha,0)\notin \psi.
\end{aligned}
\end{cases}
\end{equation}
\end{definition}

This definition formulates the resilience metric \( g_\psi \) that evaluates for a given $x \in X$ the maximum disturbance \(\varepsilon\) and the optimal parameter \(\alpha\) ensuring that the trajectories in \(\xi_x(x, \pi_\alpha,\varepsilon)\) satisfy the specification \(\psi\). The notation distinguishes between set inclusion (\(\subseteq\)) for disturbed trajectories and element membership (\(\in\)) for nominal cases because \(\xi_x(x, \pi_\alpha,\varepsilon)\) represents all possible trajectories under any disturbance sequence \(d = (d(1),\dots,d(N)) \in \Omega_\varepsilon(0)^N\), while \(\xi_x(x, \pi_\alpha,0)\) refers to one single nominal trajectory without disturbances. The value of $\varepsilon$ is equal to zero in the case where there is no controller that can lead the nominal trajectory \(\xi_x(x, \pi_\alpha,0)\) to satisfy the specification $\psi$.

\begin{remark}
 Note that, by definition, since both the dynamics $f$ and the controller $\pi_\alpha$ are continuous functions, the closed-loop system follows a continuous dynamics. Hence, when considering closed specifications, the supremum operator in the definition of resilience can be replaced by the maximum operator \cite{saoud2024}. This substitution will be adopted throughout the rest of this work.
\end{remark}

\subsection{Problem formulation}
Consider the system $\Sigma$ in \eqref{eqn:non linear controlled system}, a specification $\psi \subseteq X^{N+1}$, a controller template $\pi_\alpha$ as in Section~\ref{system definition} and a point $x \in X$. The objective is to design the optimal controller $\pi_\alpha$ such that all the trajectories satisfy the specification $\psi$, i.e., $\xi_x(x, \pi_\alpha,\varepsilon) \subseteq \psi$ while maximizing the resilience $g_{\psi}(x)$. 
 
\section{Linear systems and linear controllers}
\label{sec:3}
% This section addresses the computation of controllers maximizing the resilience metric for linear discrete-time systems. 
Consider the system $\Sigma$ in \eqref{eqn:non linear controlled system}  with a linear dynamics:
\begin{equation}
\begin{aligned}
x(k+1) 
&= Ax(k) + Bu(k) + d(k), 
\end{aligned}
\label{eqn:linear sys}
\end{equation}
where $A \in  \R^{n \times n}$ and $B \in \R^{n \times m}$  are the state and input matrices governing the dynamics. In this section, we consider a linear feedback controller defined by \( \pi_\alpha(x(k)) =  u(k) = \alpha_1 x(k) + \alpha_2 \), where $(\alpha_1, \alpha_2) \in \mathbb{R}^{m\times n} \times \mathbb{R}^{m}$.  Given a specification $\psi \subseteq X^{N+1}$, the calculation of the resilience metric requires the computation of the optimal parameters $\alpha_1$ and $\alpha_2$ that ensure the system respects the specification while simultaneously maximizing the disturbance. %The calculation of the resilience metric $g_\psi$ outputs the values of maximal disturbance $\varepsilon$ and the optimal controller parameters $\alpha_1$ and $ \alpha_2$.

In this section, we consider specifications defined by $\psi = \Gamma_0 \times \Gamma_1 \times \dots \times \Gamma_N \subseteq X^{N+1}$, where for $k=0,1,\ldots,N$, $\Gamma_k = \{x \mid G_kx \leq H_k\}$ with $G_k \in \mathbb{R}^{q \times n}$ and $H_k \in \mathbb{R}^q$. We provide a closed-form expression of the resilience metric such that the state evolution at time $k$ belongs to the set \( \Gamma_k \). Note that such specifications include the exact time reachability, finite-horizon safety and the more general class of convex LTL$_f$ specifications {(for details on convex specifications, please refer to Definition 3.3 and Section 7.3 in \cite{saoud2024}).}
In the following, we show that the computation of the resilience metric can be formulated as an optimization problem. 

\begin{theorem}
\label{theorel:linear_farkas}
    Consider the controlled discrete-time linear system $\Sigma$ in \eqref{eqn:linear sys} with a linear controller $\pi_\alpha(x(k)) = \alpha_1 x(k) + \alpha_2$. Consider the specification $\psi= \Gamma_0 \times \Gamma_1 \times \dots\times \Gamma_N \subseteq X^{N+1}$, where $\Gamma_i=\{x \in X \mid G_i x \leq H_i\}$ and $G_i \in \mathbb{R}^{q \times n}$ and $H_i \in \mathbb{R}^q$, for some $N \in \mathbb{N}$. We have
\begin{equation}
\begin{aligned}
g_\psi(x)=\max_{\varepsilon \geq 0,  \alpha_1 \in \mathbb{R}^{m \times n}, \alpha_2 \in \mathbb{R}^{m}, P \geq 0}& \{\varepsilon \geq 0 \mid  
\\ P \geq 0, P A_b=E(\alpha_1),
&\varepsilon P B_b \leq  F(x, \alpha_1, \alpha_2)\}
\end{aligned}
\end{equation}
with
\begin{equation}
\label{A_b}
A_b = \begin{bmatrix}
 \mathbb{I}_{n(N+1)} \\ 
 -\mathbb{I}_{n(N+1)}
\end{bmatrix}, \quad 
B_b = \begin{bmatrix}
\mathbf{1}_{n(N+1)} \\ 
\mathbf{1}_{n(N+1)}
\end{bmatrix}
\end{equation}
\begin{equation}
\label{E(x)}
E(\alpha_1) = 
\begin{bmatrix}
\mathbf{0}_{q \times n} & \mathbf{0}& \cdots & \mathbf{0} & \mathbf{0}\\
\mathbf{0} & G_1 & \ddots & \vdots &\mathbf{0}\\
\mathbf{0}& G_2\bar{A} & G_2 & \mathbf{0} &\mathbf{0}\\
\vdots & \vdots & \ddots & \ddots&\mathbf{0}\\
\mathbf{0} & G_N\bar{A}^{N-1} & \cdots & G_N\bar{A} & G_N
\end{bmatrix}
\end{equation}
\begin{equation}
\label{F(x)}
F(x,\alpha_1,\alpha_2) = \begin{bmatrix}
H_0 - G_0x \\
H_1 - G_1\bar{A}x - G_1B\alpha_2 \\
\vdots \\
H_N - G_N\bar{A}^Nx - G_N\sum_{i=0}^{N-1}\bar{A}^iB\alpha_2
\end{bmatrix} 
\end{equation}
and $\bar{A} = A + B\alpha_1$.
\end{theorem}

The result of the previous theorem enables the transformation of the resilience metric computation from a robust optimization problem into a deterministic polynomial optimization problem. This reformulation eliminates uncertainties, making the problem computationally tractable and solvable using well-established methods \cite{NumericalOp}. %\textcolor{blue}{In contrast, robust optimization problems often require slower, scenario-based, or duality techniques, which can lead to overly conservative solutions.}

\begin{remark}
In this section, we did not consider input constraints to avoid saturation and nonlinearities, thereby maintaining a polynomial formulation. The constraints of the optimization problem are linear in $\varepsilon$ and polynomial in $\alpha_1$ and $\alpha_2$, which preserves the equivalence when applying Farkas' Lemma \cite{Alexander1999}. The theorem extends the results in \cite{saoud2024} to a broader class of linear controlled systems governed by a linear state-feedback controller under a more general class of specifications. The next section introduces the case where we take into account input constraints and employ scenario optimization to solve the required robust optimization.
\end{remark}

\section{Nonlinear systems and controllers with general specifications}
\label{scenario_section}
This section presents a scenario optimization based approach to compute the resilience metric for nonlinear systems, nonlinear controllers, and general specifications. With this approach, we can handle state constraints defined by the specification, as well as input constraints. Although the scenario approach provides only an approximate solution to the robust optimization problem, it ensures computational feasibility for practical implementation. Moreover, it also allows providing probabilistic guarantees on the obtained result \cite{garatti2024}. We first formulate the calculation of the resilience metric as a robust optimization problem and then translate it to a scenario optimization problem through a disturbance normalization step.
 
\subsection{Robust Optimization}
Consider the system defined in \eqref{eqn:non linear controlled system}, with a constrained input set given by $U  \subseteq \R^m$. To properly state the scenario optimization problem, let us define the robust optimization problem for a given $x \in X$, and a specification $\psi \subseteq X^{N+1}$ as

\begin{equation}
g_\psi(x) = 
\begin{cases}
\begin{aligned}
\label{eqn:robust_pb}
\max_{(\varepsilon, \alpha) \in \mathbb{R}_+ \times \mathbb{R}^d} \ \varepsilon \\
\text{subject to }& (\varepsilon, \alpha) \in  \mathcal{X},
\end{aligned}
\end{cases}
\end{equation}
where the constraint set $\mathcal{X}$ is given by
\begin{equation}
\label{eqn:cont_constr}
\begin{aligned}
\mathcal{X} = \big\{ &(\varepsilon, \alpha) \in \mathbb{R}_+ \times \R^d \mid  \nonumber\text{for all }  \mathbf{d} = (d_0, \ldots, d_{N-1}) \in \Omega_{\varepsilon}(0)^N \nonumber\\
& \xi_x(x,\pi_\alpha, \varepsilon) \subseteq \psi  \text{ and } \nonumber \xi_u( x,\pi_\alpha, \textbf{d}) \in U \big\}.
\end{aligned}
\end{equation}
One way to solve this problem is to consider all the possible values of $\mathbf{d} = (d_0, d_1, \dots, d_N ) \in \Omega_{\varepsilon}(0)^N$ and look for a solution. This is not possible because of the infinite realization possibilities of $\mathbf{d}$, the optimization problem is intractable. To address this, one approach is to randomly sample the $\mathbf{d}$ and solve the problem for these samples. This approach is called the scenario approach \cite{garatti2024}.
\subsection{Scenario Optimization}

To formulate the scenario optimization problem, we define the normalized disturbance set 
$\mathcal{D} = \{\delta = (\delta_0, \dots, \delta_{N-1}) \in \Omega_1(0)^N\}$, where $\mathbf{d} = \varepsilon\delta$ represents the actual disturbance in 
$\Omega_\varepsilon(0)^N$. For a given $N \in \mathbb{N}$, the state evolution under a fixed $\delta \in \mathcal{D}$ follows the recursive relation $x_\delta(N) = f\big(f(\dots f(x, \pi_\alpha(x)) + \varepsilon \delta_0, \dots\big) + \varepsilon \delta_{N-1}.$
The corresponding trajectory $\xi(x, \pi_\alpha, \varepsilon\delta)$ is the sequence of states and controls given by
$$
\begin{aligned}
\label{trajectory}
    &\xi(x, \pi_\alpha, \varepsilon\delta) = \big\{\big(x_\delta(0), u(0)\big), \big(x_\delta(1),u(1)\big), \ldots, x_\delta(N))  \mid \\
    &x(0) = x, u(k) = \pi_\alpha(x_\delta(k)),
    x_\delta(k+1) = f(x_\delta(k) , u(k)) +\varepsilon \delta_k  \big\}.
\end{aligned}
$$
For a state $x \in X$ and disturbance $\delta \in \mathcal{D}$, we define the set of constraints as follows:
\begin{equation}
\begin{aligned}
\mathcal{X}_\delta = \big\{ (\varepsilon, \alpha)& \in \mathbb{R}_+ \times \R^d \mid \\
& \xi_x(x, \pi_\alpha, \varepsilon\delta) \in \psi ,
\xi_u( x, \pi_\alpha, \varepsilon\delta) \in  U^N \big\}.
\end{aligned}
\end{equation}
This ensures the disturbed trajectory meets the specification $\psi$ while respecting input constraints $U$. 

For \(i \in \{1, \dots, M\}\), we consider \(M\) scenarios, \(\delta_i = (\delta_0^i, \dots, \delta_{N-1}^i)\), which are taken independent and identically distributed (i.i.d) from the probability space  \((\mathcal{D}, \mathcal{F}, \mathbb{P})\), where $\mathcal{F}$ is the Borel $\sigma$-algebra on $\mathcal{D}$ and \(\mathbb{P}\) is any probability measure and \(M \in \mathbb{N}\). With these concepts in hand, the scenario optimization problem is formulated as
\begin{equation}
\begin{aligned}
\label{eqn:scenario_pb}
\max_{(\varepsilon, \alpha) \in \mathbb{R}_+ \times \mathbb{R}^d} \ \varepsilon \\
\text{subject to }& (\varepsilon, \alpha) \in \bigcap_{i=1}^{M} \mathcal{X}_{\delta_i}.
\end{aligned}
\end{equation}

This nonconvex optimization problem seeks the optimal solution denoted by $\theta_M^* = (\varepsilon_M^*, \alpha_M^*) $ that is feasible for all $M$ scenarios and can be solved with known methods such as sequential quadratic programming or interior-point techniques \cite{NumericalOp}. Furthermore, we rely on the results in \cite{garatti2024} to provide a probabilistic guarantee for the generalization of the solution $\theta_M^*$ to unseen constraint scenarios. We will formulate this guarantee in the rest of this section. 

For a fixed number of scenarios $M$, we assume that a feasible and locally optimal solution $\theta_M^* = (\varepsilon_M^*, \alpha_M^*) \in \mathbb{R}_+ \times \mathbb{R}^d$ for problem \eqref{eqn:scenario_pb} is available and define the violation probability or the risk of the solution as $V(\theta_M^*)=\mathbb{P}\left\{\delta \in \mathcal{D}: \theta_M^* \notin \mathcal{X}_\delta\right\}$. This metric measures the generalization power of the optimal solution to unseen constraints. 
{It provides the probability that a new constraint will \textbf{not} be satisfied by the solution $\theta_M^*$, indicating a violation of the considered solution. This indicates the probability with which the obtained solution can be generalized to the set of uncertainties $\mathcal{X}$. In the perfect case where $V(\theta_M^*) = 0$, the solution $\theta_M^*$ remains valid for any scenario constraint in the continuous space $\mathcal{X}$ in (\ref{eqn:cont_constr}). }
A constraint in the scenario program ~\eqref{eqn:scenario_pb}  is called support constraint if its removal modifies the optimal solution  $\theta_M^*$, and the complexity $s_M^*$ of $\theta_M^*$ is the number of such support constraints.

% it is essential to ensure that the solver used delivers repeatable results which requires using deterministic solvers. 

We define the discrete function $b(k)$ for $k=$ $0,1, \ldots, M$ as
\begin{equation}
\label{eqn:epsilon}
b(k):=1-t(k), \quad k=0,1, \ldots, M-1, \text { and } b(M)=1,
\end{equation}
where $t(k)$ is the unique solution of the following polynomial equation for a chosen confidence parameter $\beta$:
\begin{equation}
\frac{\beta}{M} \sum_{m=k}^{M-1}\binom{m}{k} t^{m-k}-\binom{M}{k} t^{M-k}=0.
\end{equation}
This metric establishes the bound on the violation probability. We can now present the main theorem in this section.
\begin{theorem}
\label{thrm:scenario_optimization}
Consider the system $\Sigma$ in \eqref{eqn:non linear controlled system}. For $x \in X$, a specification $\psi \subseteq X^{N+1}$ as defined in \ref{Sec:specfication} and an input set $U \subseteq \R^m$, let $\theta^*_M$ be the solution to the optimization problem defined in \eqref{eqn:scenario_pb} for $M \in \mathbb{N}$ number of scenarios. For any probability measure $\mathbb{P}$, for any confidence $\beta \in (0, 1]$ and with $b(k)$, $ k = 0, 1, \dots, M$ as given in \eqref{eqn:epsilon}, it holds that
\[
\mathbb{P}^M \big(V(\theta_M^*) < b(s_M^*)\big) > 1-\beta,
\]
% where \( s_M^* \) is called the complexity of the solution and refers to the number of scenarios \( \delta_i \) for \( i \in \{1, \dots, M\} \) that alter the solution \( \theta_M^* \) if removed.
\end{theorem}

\medskip
{For more details on the metric $s_M^*$, called complexity, please refer to \cite{RiskComplexity}.  In addition, the information on the number of scenarios $M$ sufficient to achieve a level of confidence is given implicitly by Theorem~\ref{thrm:scenario_optimization}. }

    After fixing a confidence level $\beta$, there always exists $M$ sufficiently large such that equation (\ref{eqn:epsilon}) holds for a bound $b(k)$. Hence, one may choose the higher value of $M$ until reaching the desired violation probability bound level. This shows the fact that any confidence parameter and bound can be achieved by an appropriate choice of the number of scenarios $M$.

\section{Case studies}
\label{sec:4}
\subsection{Mobile Robot}
\begin{figure}[htbp]
    \centering
    \includegraphics[scale=0.48]{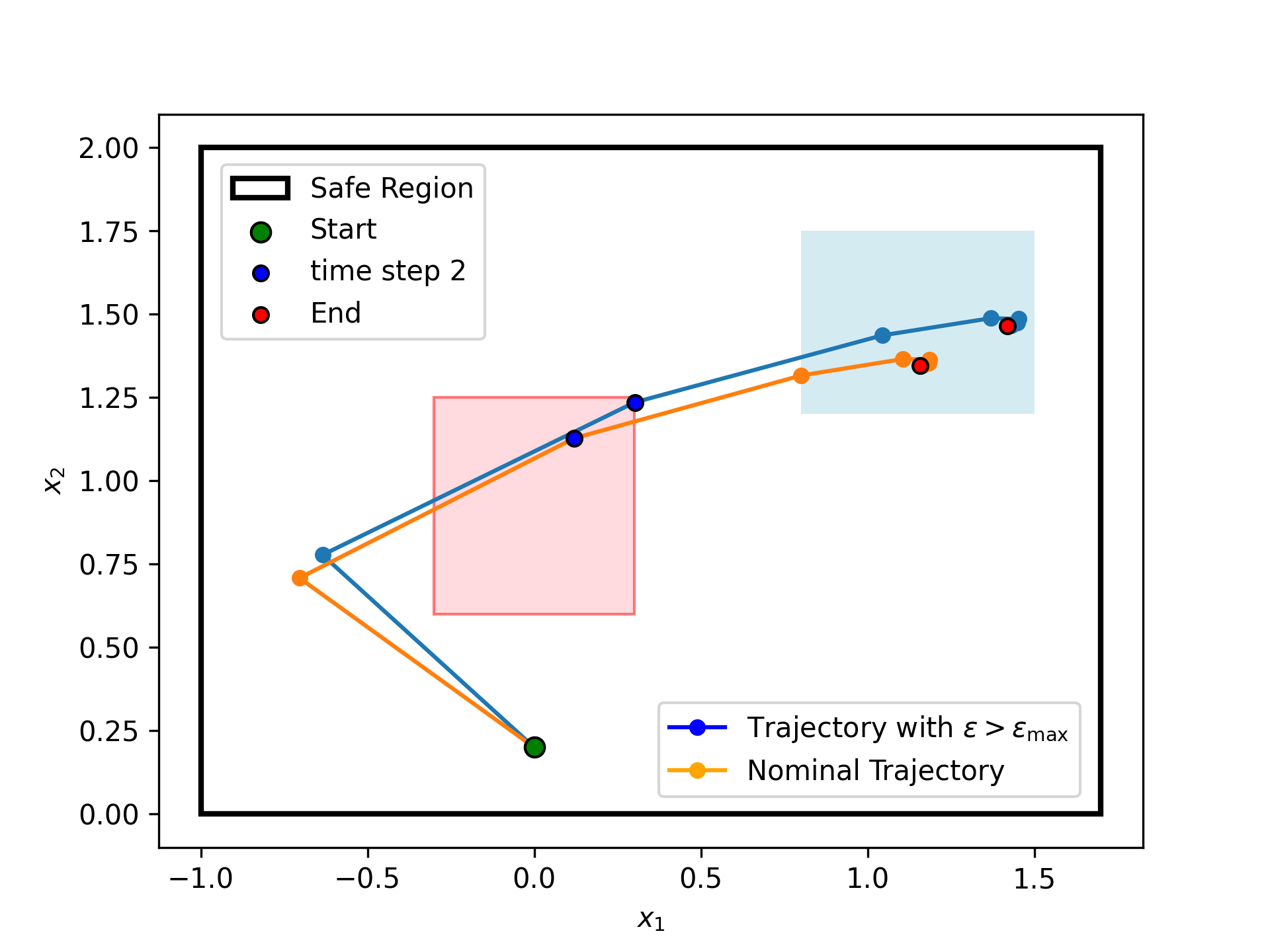}
    \caption{Illustration of two controlled trajectories of the robot starting from the initial state $x(0)=(0, 0.2)$ for $6$ time steps where both trajectories must satisfy the specification $\psi$ in Equation \eqref{spec:psi}. The yellow trajectory represents the state evolution without disturbances $\varepsilon = 0$ while the blue trajectory represents the evolution with a disturbance exceeding resilience bounds $\varepsilon > g_\psi(x(0))$.}
    \label{fig:trajectory_nominal_max_dist}
\end{figure}

We consider the linear dynamics of a mobile robot as in \eqref{eqn:linear sys}, where $A = \mathbb{I}_2\in  \R^{2 \times 2}$ and $B =\mathbb{I}_2 \in \R^{2 \times 2}$. The input is controlled using the linear controller $\pi_\alpha(x(k)) = u(k) = \alpha_1 x(k) + \alpha_2$, where $(\alpha_1, \alpha_2) \in \mathbb{R}^{2\times 2} \times \mathbb{R}^{2}$ are the controller parameters. 
The state is a two-dimensional vector characterized by the position $x \in\left[-1, 1.7\right] \times\left[0, 2\right]$ and the input vector $u \in  \R^2$ without constraints.
To describe the desired behavior of the system, we first define the three regions: $R_1= [-0.3, 0.3] \times [0.6, 1.25]$, $R_2 = [0.8, 1.5] \times  [ 1.2, 1.75]$ and $R_3 = [-1, 1.7] \times [0, 2]$ as shown in Figure~\ref{fig:trajectory_nominal_max_dist}. The desired behavior we want to satisfy is given by the LTL formula
\begin{equation}
\label{spec:psi}
    \psi = \nex^2 R_1 \wedge  \square^{[4,6]} R_2 \wedge \square^{[0,6]} R_3, 
\end{equation}
which can be described as follows: remain in region $R_3$ from the start until step $6$ ( $\square^{[0,6]} R_3$), while reaching region $R_1$ at time step $2$ ($\nex^2 R_1$) and staying in region $R_2$ between time instances $4$ and $6$ ($\square^{[4,6]} R_2$).

All trajectories in this example start from the same initial condition $x(0) = (0, 0.2)$. After solving the optimization problem outlined in Theorem~\ref{theorel:linear_farkas}, we obtain the following values for the controller parameters: 
$$
\alpha_1 = \begin{pmatrix} -0.99 & 1.62 \\ -0.11 & -0.33 \end{pmatrix}, \quad \alpha_2 = (-1.028, 0.574),
$$
and the corresponding maximum disturbance is $g_{\psi}(x(0)) = 0.0686$. These values were obtained by solving the optimization problem using the Python library Pyomo and the Ipopt solver, which is implemented based on the {interior-point methods} \cite{NumericalOp}. 

\begin{figure}[htbp]
    \centering
    \includegraphics[scale=0.55]{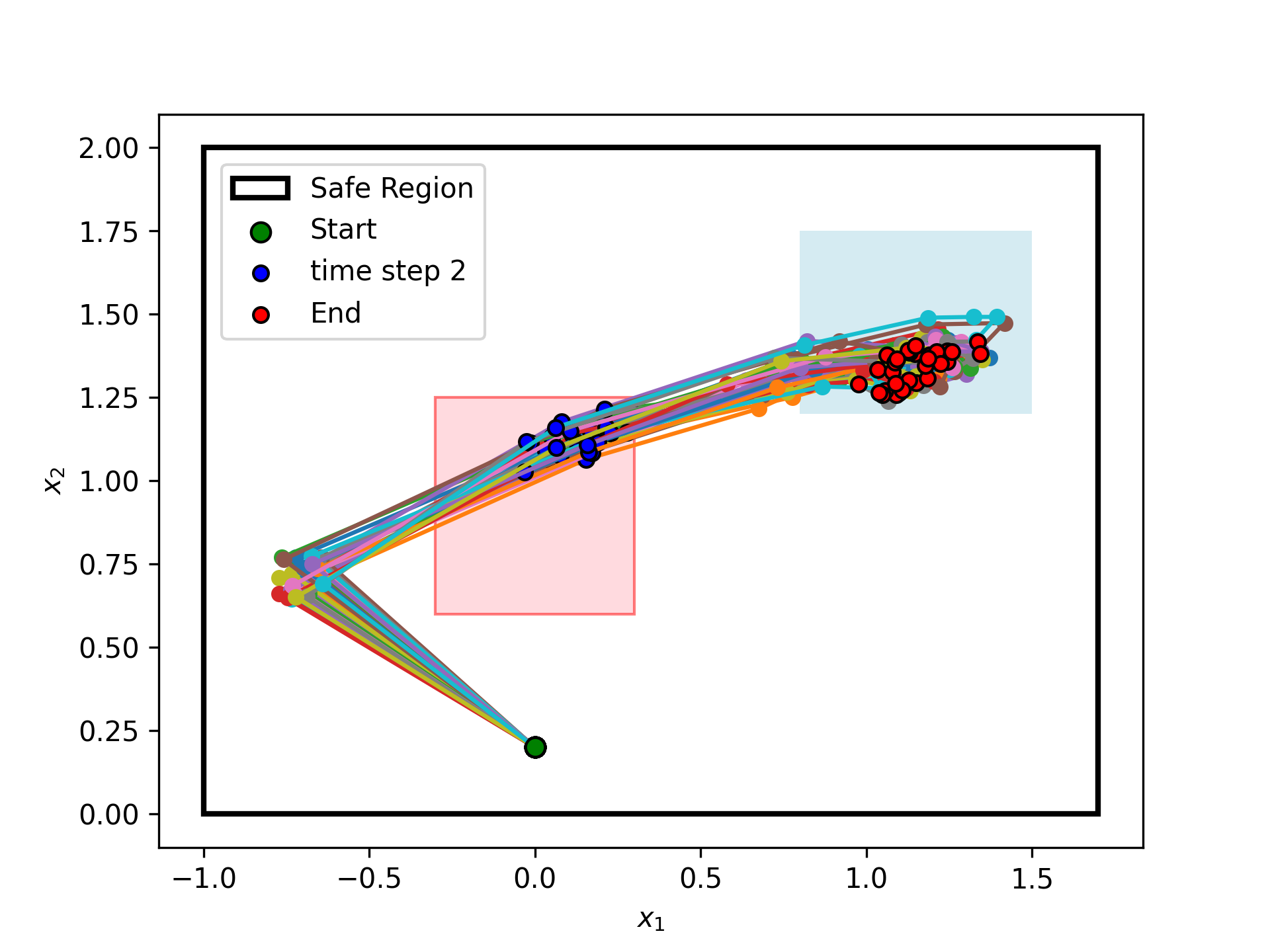}
    \caption{Illustration of controlled trajectories of the robot starting from the initial state $x(0)=(0, 0.2)$ for $6$ time steps under the specification $\psi$ in Equation \eqref{spec:psi} with disturbances within the resilience bounds $\varepsilon \leq g_{\psi}(x(0))$.}
    \label{fig:trajectory_within_disturbance}
\end{figure}

We demonstrate how the system behaves under the optimal controller and with three different cases of disturbance. First, the resilient case in Figure~\ref{fig:trajectory_within_disturbance} shows how the system respects the specification when facing $100$ sampled disturbances less than or equal to the value $g_{\psi}(x(0)) = 0.0686$ given by the resilience metric in Equation \eqref{eq:resilience} where each disturbance corresponds to a trajectory.  We can observe that the specification is satisfied: all the trajectories are inside the region $R_3$, they reach region $R_1$ (in light red) at time step $2$, and they reach and remain in region $R_2$ (in light blue) between time instances $4$ and $6$. 
\begin{table}[H]
\centering
\setlength{\tabcolsep}{3pt} % Tight column spacing
\begin{tabular}{|c|c|c|c|c|c|c|}
\hline
Parameter & $m$ & $f_0$ & $f_1$ & $f_2$ & $\underline{F}$ & $\overline{F}$ \\ \hline
Value & 1370 & 51.0709 & 0.3494 & 0.4161 & -4031.9 & 2687.9 \\ \hline
Unit & kg & N & Ns/m & Ns²/m² & kg·m/s² & kg·m/s² \\ \hline
\end{tabular}
\caption{Parameters for the adaptive cruise control system in Section~\ref{ACC}}
\label{fig:ACC_values}
\end{table}

Figure~\ref{fig:trajectory_nominal_max_dist} shows two cases: the yellow trajectory of the system which corresponds to a nominal trajectory without any disturbance, and the violation case when disturbances exceed the resilience value $\varepsilon > g_{\psi}(x(0))  = 0.0686$, highlighting the violation of the specification. As we can notice, the state at time step 2 in the blue trajectory is not inside the region $R_1$. 
The results in the figures confirm the control objective is achieved using the resilience metric values by solving the optimization problem in Theorem~\ref{theorel:linear_farkas} which represents the maximal disturbance for which the trajectories $\xi_x(x(0), \pi_\alpha, g_{\psi}(x(0))) $ satisfy the specification $\psi$. The sharp transition from satisfaction to violation at this threshold demonstrates the tightness of our result.

\subsection{Adaptive Cruise Control} 
\label{ACC}
\begin{figure}
     \centering
     \includegraphics[width=0.48\textwidth]{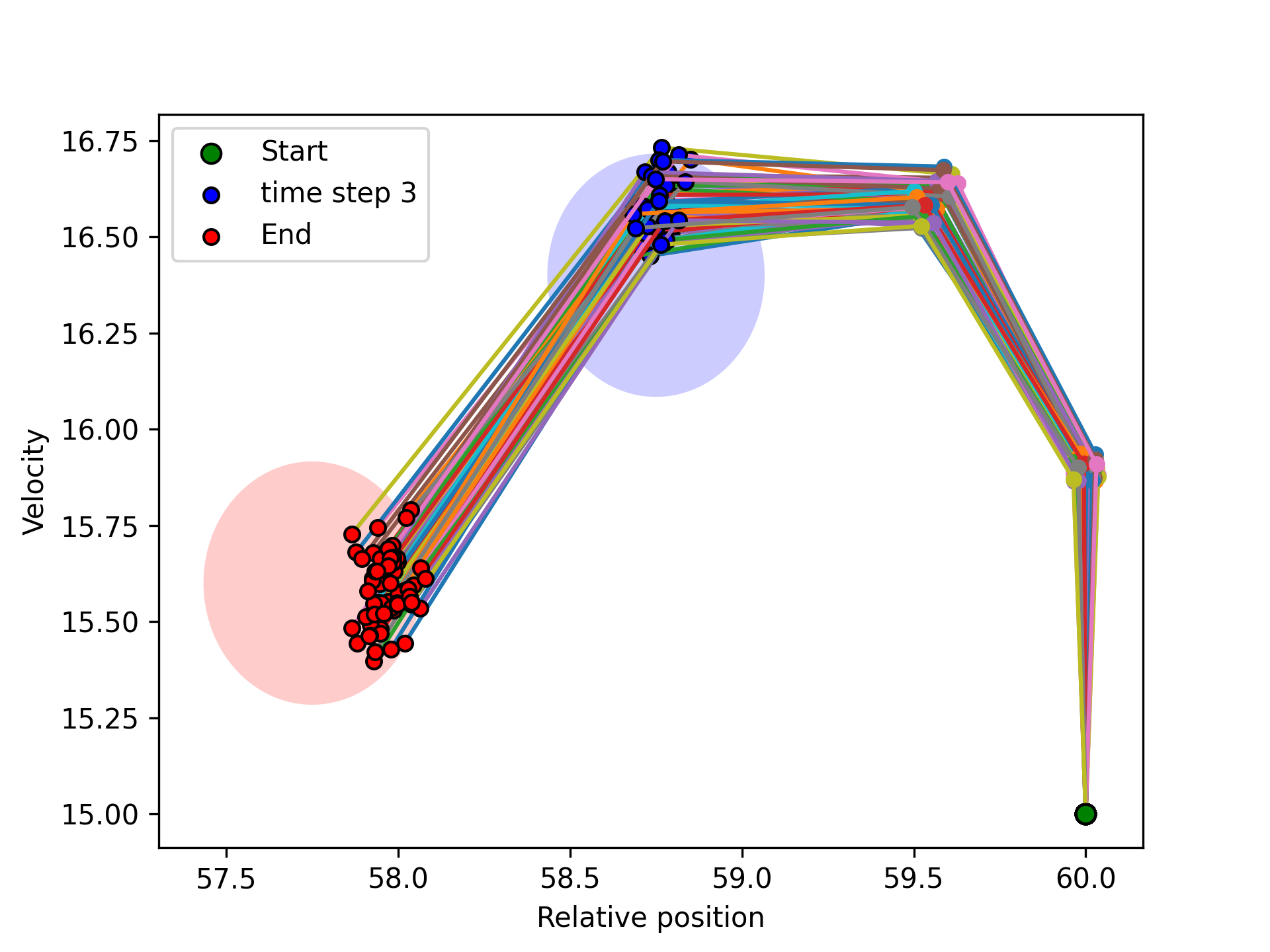}
     \caption{Illustration of Adaptive Cruise Control trajectories with the optimal linear controller and with disturbances less than the maximum disturbance given by the resilience metric $\varepsilon \leq g_{\psi}(x(0))=0.03 $ found for $M =100$ scenarios.}
\label{fig:linear_ACC}   
\end{figure}

Adaptive Cruise Control (ACC) is a classic example of controlled dynamical systems in vehicles. Consider a vehicle following a lead car moving at a constant velocity $v_0$ on a straight road. The following vehicle adjusts its speed to maintain a safe distance while responding to changes in the environment or road conditions. The vehicle dynamics, adapted from \cite{ACC_Saoud}, are described by the differential equation
$$
\left\{\begin{aligned}
h(k+1) & =h(k)+\tau\left( \delta v_0 + v_0-v(k)\right) \\
v(k+1) & =v(k)+\frac{\tau}{m}\left(F(k)+ \delta f_0 -f_0-f_1 v(k)-f_2 v(k)^2\right),
\end{aligned}\right.
$$
where $v \geq 0$ represents the velocity of the vehicle, $h$ the distance between the lead and second vehicle, $m>0$ is the mass of the vehicle,  and the term $f_0+f_1+f_2 v^2$ includes the rolling resistance and aerodynamics and $\tau$ represents a sampling period. The disturbance is modeled by $\delta v_0$, which is the uncertainty on the velocity $v_0$ of the lead vehicle and $\delta f_0$ is the uncertainty on the parameter
$f_0$.
The variable $F$ represents the control input and must satisfy $F \in [\underline{F}, \overline{F}]$ for values given in Table~\ref{fig:ACC_values} along with the constants of the dynamics borrowed from \cite{ACC_Saoud}. All trajectories in this example start from the same initial condition $x(0) = (60, 15)$. We have conducted two experiments, the first using a linear controller defined for $x =(h, v) \in \R^2$ as $\pi_\alpha(x) = \alpha_1 x + \alpha_2$, where $\alpha_1 \in \R^2$ and $\alpha_2 \in \R$ are the parameters of the controller.
In this experiment, we forced the behavior expressed by the following specification: $\psi=\psi_1 \wedge \psi_2$ with $\psi_1:=\bigcirc^3B_1$ and $\psi_2:=\nex^4 B_2$. This behavior can be interpreted as follows: the relative position and velocity should reach the set $B_1= \mathcal{B}_{\sqrt{0.1}}((58.75, 16.4))$ in $3$ steps and reach the set $B_2= \mathcal{B}_{\sqrt{0.1}}((57.75, 15.6))$ in $4$ steps, which means that we will force the velocity to increase which result in the decrease of the distance between the two cars.
The objective is to compute the resilience metric under which the trajectory of the system initiated from $x(0)=(60,15)$ satisfies the specification $\psi$.  
The dynamical system in this use case is non-linear, and we used a scenario approach described in the Section~\ref{scenario_section}. The solution of the optimization defined by the resilience metric $g_\psi$ was obtained for a set of i.i.d. scenarios $M$ sampled using uniform probability measure on the space of disturbances $\mathcal{D}$. This optimization was performed using the Python library Pyomo and solved with the Ipopt solver, which is based on interior-point methods \cite{NumericalOp} that takes 4s to converg.
The resulting values of the optimization and the complexity $s_M^*$  for different values of scenarios $M$, are given in Table~\ref{tab:scnario_table}, as well as the values of the violation bound $b$ for different values of the confidence parameter $\beta$. 
One can see that increasing the number of scenarios decays the value of the disturbance, which is expected since exploring more scenarios tends to include more constraints and allows for a tighter approximation of the resilience metric. Figure~\ref{fig:linear_ACC} shows trajectories with disturbance less than the resilience metric $g_{\psi}(x(0)) = 0.036$ found for $M = 100$. {In this scenario, the computation of the controller maximizing the resilience metric takes 4 seconds}. One can readily see that the controlled trajectories reach the target set $B_1$ (in light blue) in three time steps and then reach the target set $B_2$ (in light red) in four time steps. At four time steps, we observe that only few points fall outside the set $B_2$, which confirms that, while our solution is not exact, it is probabilistically guaranteed. The values of the variable $\delta v_0$ and $\delta f_0$ for maximal resilience metric results correspond to the following intervals $v_0 = [-0.072, 0.072]$ 
and $f_0 = [-98.36,98.36]$.
%
%$f_0 = [−82.2, 82.2]$
%and $f_0 = [−82.2, 82.2]$.
%
\begin{table}
\centering
\resizebox{0.485\textwidth}{!}{% %
\begin{tabular}{|c|c|c|c|}
\hline
\diagbox[width=3cm, height=1.2cm]{variables}{ scenarios (M)}& $10 $ & $100$ & $500$\\
\hline
$\varepsilon^*_M$ & $0.039$& $0.0367$& $0.0305$\\
\hline
$\alpha_1^*$ &  $[23428.089 ,-567.85]$&   $[3377.689, -599.61]$& $[3426.05,-588.90] $\\
\hline
$\alpha_2^*$ &  $-194479.59$&  $-190979.28$&   $ -194041.69$\\
\hline
$s_M^*$ &  $4$&  $8$&   $9$\\
\hline
$b(s_M^*, \beta = 10^{-2})$&  $0.851$&  $0.202$& $0.046$\\
\hline
$b(s_M^*, \beta = 10^{-4})$& $0.936$& $0.259$& $0.059$\\
\hline
$b(s_M^*, \beta = 10^{-6})$& $0.971$& $0.307$& $0.072$\\
\hline
\end{tabular}}
\caption{Results of the scenario optimization defined in \eqref{eqn:scenario_pb} for different values of the number of scenarios $M$, as well as the values for the bound $b$ of the risk for different values of complexity $s_M^*$ and $\beta$. }
\label{tab:scnario_table}
\end{table}

To explain the values in Table~\ref{tab:scnario_table}, let us take the column of the number of scenarios $M = 500$. The corresponding value of the complexity $s^*_M = 9$, which corresponds to the scenarios having an impact on the optimization problem. Using these $9$ scenario constraints instead of the $500$ constraint scenarios in the optimization problem, we can construct the same resilience metric solution $g_\psi(x(0))$. 
Then, to calculate the bound, we choose a level of confidence $\beta$. Taking $ \beta = 10^{-2}$, we can now calculate the bound that corresponds to $ b(s_M^*) = 0.046$. Hence, the probability of violation is bounded by $0.046$ with confidence $1-10^{-2}$. 
The expression defined in Theorem~\ref{thrm:scenario_optimization} as 
$
\mathbb{P}^M \big(V(\theta_M^*) < b(s_M^*)\big) > 1 - 10^{-2} = 0.99
$,
means that we are requiring a $99\%$ confidence level that the violation probability is below the value \( b(s_M^*) \). In this case, there is only a $4.6\%$ chance that a new scenario constraint will not be satisfied by the solution \( \theta_M^* \).
The relationship between the bound \( b \) and the confidence level \( \beta \) is inversely proportional, as demonstrated in the table. For \( M = 10 \), the bound on the violation probability is very large \( 0.851 \). However, as we increase the number of scenarios to \( M = 500 \), the bound on the violation probability decreases, reaching a reasonable value equal to \( 0.046 \). This indicates that by increasing the number of scenarios, we are able to obtain a tighter and more reliable bound on the violation probability, effectively reducing the likelihood of a scenario violating the solution. 
{To illustrate the case of nonlinear controllers, we have defined a polynomial controller of degree $2$ defined for $(h, v) \in \R^2$ as $\pi_\alpha(h, v) = \alpha_1 h^2 + \alpha_2 v^2 + \alpha_3 h v + \alpha_4 h + \alpha_5 v+ \alpha_6$. We want the system to satisfy the same specification $\psi$ used for linear controllers.
The resulting controlled trajectories in Figure~\ref{fig:polynomial_ACC} reach the target set $B_1$ (in light blue) in three time steps and reach the target set $B_2$ (in light red) in four time steps except for a small number of states, as discussed in the previous example. The resulting value of the resilience metric for $M = 100$ is $g_{\psi}(x(0)) = 0.078$, with controller parameters given by $\boldsymbol{\alpha}  =
[-501.63,\; -997.46,\; 2142.48,\; 27197.67, \\
\quad -97858.42,\; -59225.81]^\top
.$

\section{Conclusion and future work}
\label{sec:conclusion}
We provided a resilience metric framework for designing controllers for dynamical systems. For a given finite specification that defines the desired behavior of the system’s trajectory, this metric quantifies the optimal controller ensuring the system satisfies the specification and the maximum disturbance for which the specification remain respected. For linear dynamical systems and linear controllers, we demonstrated using the Farkas' lemma how to compute the exact resilience metric. In the general case of nonlinear systems and nonlinear controllers, a scenario approach applied to nonconvex optimization problems was used to derive the resilience metric while having a probabilistic guarantee on the solution. In future work, we aim to explore how the resilience metric can be extended to interconnected controlled systems and continuous-time controlled dynamical systems. Additionally, we plan to develop algorithms that enable faster computation of the resilience metric to improve efficiency.

\begin{figure}[t]
     \centering
     \includegraphics[width=0.48\textwidth]{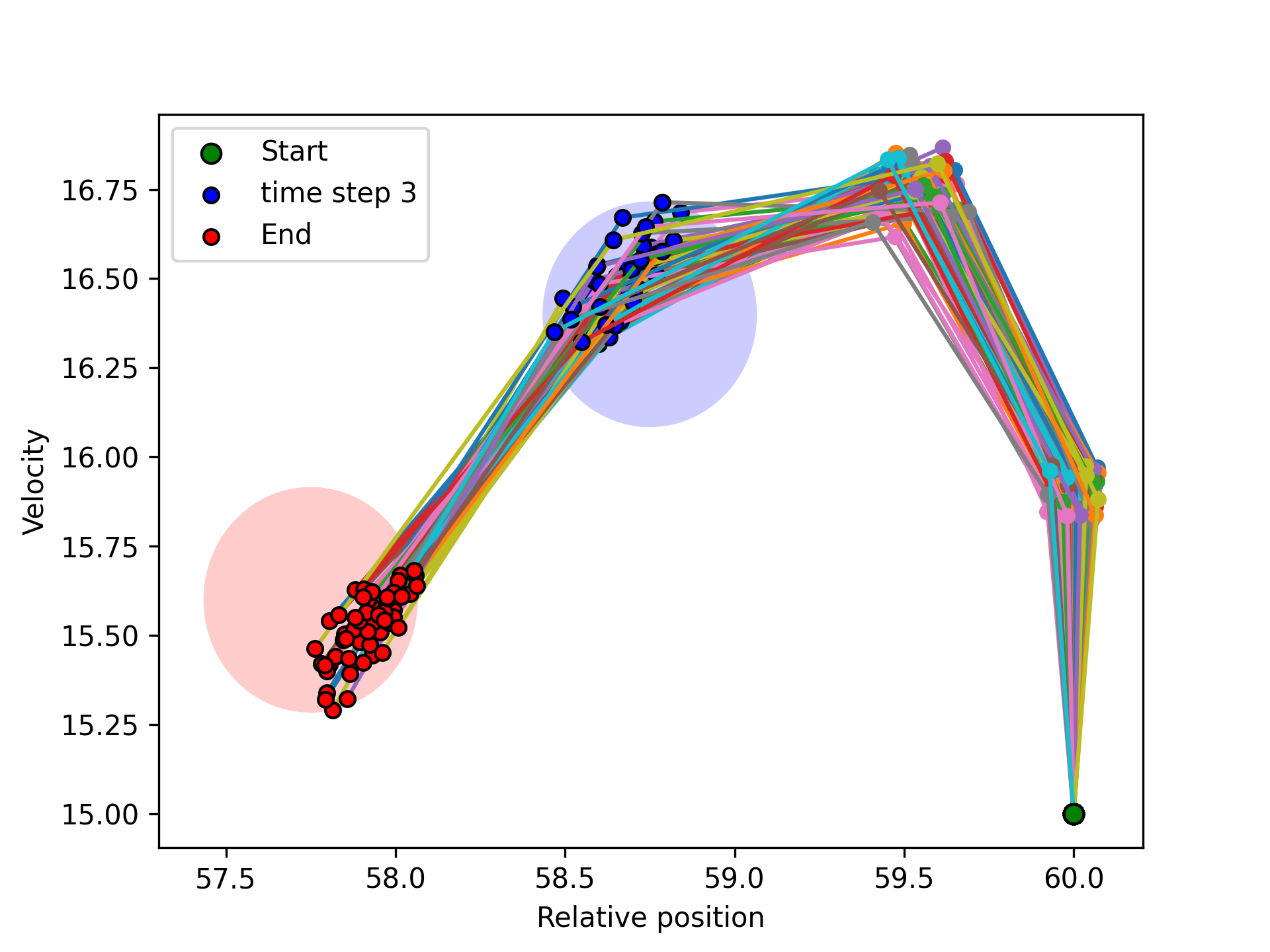}
     \caption{ Illustration of adaptive cruise control trajectories with the optimal polynomial controller and with disturbances less than the maximum disturbance given by the resilience metric $\varepsilon \leq g_{\psi}(x(0))=0.045 $ found for $M =100$ scenarios.}
\label{fig:polynomial_ACC}
\end{figure}

\bibliographystyle{IEEEtran}  % For IEEE style
\bibliography{references}     % Assumes your .bib file is "references.bib"
\appendix
\subsection{Proofs}

\subsubsection{Proof of theorem~\ref{theorel:linear_farkas}}
The system evolves according to
\begin{align*}
 x(k+1) &= Ax(k) + Bu(k) + d(k)\\
 & = \bar{A}x(k) + B\alpha_2 + d(k)
\end{align*}
with $\bar{A} = A + B\alpha_1$. By denoting $x = x(0)$, the state at time $k \geq 1 $ is
\begin{equation}
x(k) = \bar{A}^k x + (\sum_{i = 0}^{k - 1} \bar{A}^i) B \alpha_2 + \sum_{i = 0}^{k - 1} \bar{A}^{k-1-i}d_i.
\end{equation}
Using this expression, the definition of the resilience in \eqref{eq:resilience} can be written as
$$
\begin{aligned}
g_\psi(x)= & \max_{\varepsilon \geq 0,  \alpha_1 \in \mathbb{R}^{m \times n}, \alpha_2 \in \mathbb{R}^{m}} \varepsilon \geq 0 \mid \text {for all } d_0, \ldots, d_{N-1} \in \Omega_{\varepsilon}(0) \\
& \left\{\begin{array}{l}
G_0 x \leq H_0 \\
G_1\left(\bar{A} x+B \alpha_2 +d_0\right) \leq H_1 \\
G_2\left(\bar{A}^2 x+(I + \bar{A})B \alpha_2 + \bar{A} d_0+d_1\right) \leq H_2 \\
\vdots \\
G_N\left(\bar{A}^N x+(\sum_{i = 0}^{N - 1} \bar{A}^i) B \alpha_2 + \sum_{i = 0}^{N - 1} \bar{A}^{N-1-i}d_i\right) \leq H_N.
\end{array}\right.
\end{aligned}
$$
Which is equivalent to
$$
\begin{aligned}
g_\psi(x)= & \max_{\varepsilon \geq 0,  \alpha_1 \in \mathbb{R}^{m \times n}, \alpha_2 \in \mathbb{R}^{m}} \varepsilon \geq 0 \mid \text {for all } d_0, \ldots, d_{N-1} \in \Omega_{\varepsilon}(0) \\
& \left\{\begin{array}{l}
0\leq H_0 - G_0 x  \\
G_1 d_0 \leq H_1 -G_1 \bar{A} x-G_1B \alpha_2 \\
\vdots \\
\sum_{i = 0}^{N - 1} \bar{A}^{N-1-i}G_Nd_i \leq H_N - G_N\bar{A}^N x\\ -G_N(\sum_{i = 0}^{N - 1} \bar{A}^i) B \alpha_2.  
\end{array}\right.
\end{aligned}
$$
Therefore, the computation of the resilience metric $g_\psi$ is equivalent to solving the optimization problem
\begin{equation}
\label{eqn:g_psi_simplified}
\begin{aligned}
\max_{\varepsilon \geq 0, \alpha_1 \in \mathbb{R}^{m \times n}, \alpha_2 \in \mathbb{R}^{m}} &\Big\{ \varepsilon \geq 0 \mid \, E(\alpha_1) Y \leq \frac{1}{\varepsilon} F(x, \alpha_1, \alpha_2), \\
& \text{for all } Y \text{ satisfying } A_b Y \leq B_b \Big\},
\end{aligned}
\end{equation}
with $Y:=\frac{1}{\varepsilon}\left[0_n, d_0^T, d_1^T, \ldots, d_{N-1}^T\right]^T$ being free variables, and $F(x, \alpha_1, \alpha_2)$, $ E(\alpha_1)$, $A_b$ and $B_b$ as defined in  \eqref{A_b}, \eqref{E(x)} and \eqref{F(x)} respectively. With these matrices, we have $A_b Y \leq B_b$ representing the inequality $\|Y\|_{\infty} \leq 1$ in matrix form.
Let us define for $\alpha_1 \in \mathbb{R}^{m \times n}$ and $ \alpha_2 \in \mathbb{R}^{m}$
\begin{equation}
\begin{aligned}
\varepsilon^*(\alpha_1, \alpha_2)& =\max_{\varepsilon \geq 0,}\Big\{ \varepsilon \geq 0 \mid  \, E(\alpha_1) Y \leq \frac{1}{\varepsilon} F(x, \alpha_1, \alpha_2), \\
& \text{for all } Y \text{ satisfying } A_b Y \leq B_b \Big\}.
\end{aligned}
\end{equation}
The optimization problem in \eqref{eqn:g_psi_simplified} is equivalent to
\begin{equation}
\begin{aligned}
\max_{ \alpha_1 \in \mathbb{R}^{m \times n}, \alpha_2 \in \mathbb{R}^{m}} \varepsilon^*(\alpha_1, \alpha_2).
\end{aligned}
\end{equation}
Hence, using the affine form of Farkas Lemma \cite{Alexander1999} (see Theorem~\ref{theorem farkas} in Section~\ref{Sec:aux}) we deduce that
\begin{equation}
\begin{aligned}
 \varepsilon^*(\alpha_1, \alpha_2)=\max_{\varepsilon \geq 0}  \{\varepsilon \geq 0 \mid &
P \geq 0, P A_b=E(\alpha_1),\\
&P B_b \leq \frac{1}{\varepsilon} F(x, \alpha_1, \alpha_2) \}.
\end{aligned}
\end{equation}
Finally, we conclude that
\begin{equation}
\begin{aligned}
g_\psi(x)&=\max_{\varepsilon \geq 0,  \alpha_1 \in \mathbb{R}^{m \times n}, \alpha_2 \in \mathbb{R}^{m}}  \{\varepsilon \geq 0 \mid\\ 
&P \geq 0, P A_b=E(\alpha_1),
 \varepsilon P B_b \leq  F(x, \alpha_1, \alpha_2) \},
\end{aligned}
\end{equation}
which proves the required result.\hfill\fbox{}

\subsubsection{Proof of theorem~\ref{thrm:scenario_optimization}}
 To derive formal guarantees from solving problems in \eqref{eqn:scenario_pb}, we must satisfy a requirement formalized in Property $1$ in \cite{garatti2024}. This property has been proven to hold for robust optimization problems in \cite[Section 3]{garatti2024}. Using a deterministic solver and consistent initialization, we ensure that the solver produces repeatable results for the same set of scenarios. Therefore, this property is satisfied in the case of the optimization problem defined in \eqref{eqn:scenario_pb}. This property replaces the non-degeneracy assumption introduced in \cite{wait&judge}, which is generally assumed for convex problems but does not apply to nonconvex problems. Thus, using \cite[Theorem 6]{garatti2024} for a robust optimization problem provides the desired result. \hfill\fbox{}

\subsection{Auxilliary results}
\label{Sec:aux}

The following theorem is a simple adaptation of the result in \cite[Corollary 7.lh]{Alexander1999}, and is known as the affine form of Farkas' lemma.
\begin{theorem}
\label{theorem farkas}
Suppose the set $\{x\mid E x \leq F\}$ is not empty. The following two statements are equivalent:
\begin{itemize}
    \item $Ex \leq F$ holds for all $x$ with $A x \leq b$;
    \item There exists a non-negative matrix $P$ such that $P A=E$ and $P b \leq F$.
\end{itemize}
\end{theorem}
\end{document}